Contribution to the history of astrometry No. 6          6 July 2017

# Selected astrometric catalogues

*Erik Høg, Niels Bohr Institute, Copenhagen, Denmark*

ABSTRACT: A selection of astrometric catalogues are presented in three tables for respectively positions, proper motions and trigonometric parallaxes. The tables contain characteristics of each catalogue showing the evolution in optical astrometry, in fact the evolution during the past 2000 years for positions. The number of stars and the accuracy are summarized by the weight of a catalogue, proportional with the number of stars and the statistical weight. The present report originally from 2008 was revised in 2017 with much new information about the accuracy of catalogues before 1800 AD. For the ongoing Gaia mission the website and Høg (2017) may be consulted.

## Introduction

The 400 years of astrometry from Tycho Brahe to the Hipparcos mission have been reviewed (Høg 2008d) for the symposium held at ESTEC in September 2008 to celebrate the 400 years of astronomical telescopes. For this purpose the Tables 1 to 3 were elaborated, containing data for *selected astrometric catalogues* for positions, proper motions and trigonometric parallaxes, respectively. The tables give characteristics of each catalogue showing the evolution over the past 400 years in optical astrometry. The number of stars, $N$, and the accuracy, i.e. the standard error, $s$, are summarized by the weight of a catalogue, $W$, defined in all tables as $W = N \, s^{-2} \, 10^{-6}$, proportional with the number of stars and the statistical weight. The table entries are documented in a separate paper (Høg 2017) where the origin of the new information on catalogues before 1800 AD is given. A large list of 2087 catalogues has been published by Sevarlic et al. (1978).

## Position catalogues, Table 1

An approximate standard error of individual mean position coordinates at the mean epoch of the catalogue is given in Table 1, preferably the median value as representative for the bulk of stars in a catalogue. The value $s = 0.04$ arcsec is adopted for FK5 at the mean epoch as we have derived from the comparison with Hipparcos by Mignard & Froeschlé (2000), and we assume that PPM is the only catalogue before Hipparcos which has been more accurate than 0.1 arcsec. The weight is only derived for observation catalogues, not for compiled ones. Note that position catalogues after 1990 already appear a few years after ending the observations thanks to modern computing facilities, to the available reference systems provided by Hipparcos, and to publication on the web. Previously, the reduction and publication on paper could take decades.

Some important effects are not expressed in the tables such as the serious difference in accuracy between the northern and southern celestial spheres for ground-based catalogues before 1997, after which time a N-S effect is absent due to the use of Hipparcos all-sky results. The variation within a catalogue of the accuracy with magnitude of the star, and with the observational history also does not appear from the tables. The tables give only an imperfect impression of the enormous efforts by astrometrists during the centuries. The following information is obtained from Knobel (1877), Hamel (2002), Eichhorn (1974), and Verbunt & van Gent (2010, 2012), from correspondence with colleagues and the web. No attempt for completeness is made and more details can be found in Høg (2008c).

**Positions from quadrants and sextants:** The catalogues with 1000 stars from Ptolemy, Ulugh Beg and Tycho Brahe were the largest up to Hevelius' catalogue of 1690. The catalogue of 383 stars observed in  Kassel was ready in 1587. These four catalogues are described in Høg (2017).

In 2008 we believed that Hevelius' catalogue from 1690 had three times smaller errors than Tycho's, in 2016 we know it has nearly the same accuracy but fewer outliers. Flamsteed's *Historia Coelestis Britannica* with 2935 stars, published in 1725, remained the largest until it was surpassed by N.L. de Lacaille's catalogue of 10,000 southern stars, observed from the Cape of Good Hope about 1752. Lalande's much larger *Histoire Céleste Francaise* with 50,000 stars was published in 1801. This catalogue was so important that it was republished in England by Baily half a century later, compiled with other observations. Bradley's observations beginning in 1743 were processed by the young Bessel, and again by Auwers, more than a hundred years later, resulting in positions with an accuracy of 1.1 arcsec as mentioned by Høg (2008d).

**Positions from visual meridian circles:** The Geschichte des Fixsternhimmels (GFH) contains the observations of 365,000 stars from 492 catalogues before 1900 compiled after 1899 by F. Ristenpart and many others, and published in 48 volumes between 1922 and 1964. The GFH was supplemented by Index der Sternörter I and II, initiated by R. Schorr in 1924. It gives reference to 401 catalogues of the observations after 1900 of 365,000 stars, half of them from the southern hemisphere(!), and was  published 1927-1966.

**The instruments in Col. 2 of Table 1 and 2:** Observations were visual before 1880 and with all meridian circles in the list, except where noted. Abbreviations: *AQua* = azimuth quadrant, introduced in Europe by the Landgrave Wilhelm IV in Kassel in 1558, *Sext* = sextant, invented by Tycho Brahe in  1570,  *MC* = meridian circle, *AAC* = alt-azimuth circle, *TI* = transit instrument, *Pgr.* = photographic, *p.e.* = photoelectric, and *sat.* = satellite.

**Table 1b**   From the **now obsolete Table 1 of 2008**: Position catalogues before 1700 AD.  The instruments used by Ptolemy and Ulugh Beg were listed in 2008 as *Sextant* which is certainly incorrect because this instrument was invented by Tycho Brahe in 1570. The number of stars in Wilhelm's catalogue is 383, not 1004 as often told in the literature.

| Catalogue | Instrument | Publ. | Mean epoch | Obs. period | N | $s_{star}$ |
|---|---|---|---|---|---|---|
| | | *year* | *year* | *years* | *entries* | s.e. of star *arcsec* |
| Ptolemy | Sextant | 150 | 138 | | 1025 | 1 deg. |
| Ulugh Beg | Sextant | 1665 | 1437 | 17 | 1018 | 1 deg. |
| Wilhelm of Hesse | Quad | 1594 | | | 1004 | 360" |
| Tycho Brahe | Many | 1598 | 1586 | 20 | 1005 | 60" |
| Hevelius | Quad+Sext | 1690 | 1670 | 20? | 1564 | 20" |

**Table 1**  Position catalogues as of 2017.  The standard error of the mean positions in a catalogue is a median value, if available, representing the bulk of stars, mostly faint ones. The standard errors in the three tables are often internal errors only, e.g. those in brackets (...); the external errors may be much larger, cf. Høg (2008c).  - **Ulugh Beg** observed about 1437 and his catalogue was soon known to muslim astronomers, but the first Western edition of his catalogue is from 1665. The two years of publication of **Lacaille's** catalogue are given in Høg (2017).  **Argelander's** accurate meridian circle catalogue in the list is *not* the same author's BD survey catalogue with approximate positions of 325,000 stars.

| Catalogue | Instrument | Publ. | Mean epoch | Obs. period | N | n | $s_{star}$ s.e. of star | W weight |
|---|---|---|---|---|---|---|---|---|
| | | *year* | *year* | *years* | *entries* | *per star* | *' or "* | |
| Ptolemy | - | c.150 | 138 | | 1028 | | 25' | |
| Ulugh Beg | - | c. 1437 | 1437 | 17 | 1018 | | 20' | |
| Wilhelm in Kassel | AQua, Sex | 1587 | 1586 | 3 | 384 | | 1.14' | |
| Tycho Brahe | Many | 1598 | 1586 | 20 | 1004 | | 2.0' | 0.000,000,07 |
| Hevelius | Quad+Sext | 1690 | 1670 | 20? | 1564 | | 2.0' | 0.000,000,11 |
| Rømer | MC | (1735) | 1706 | 0.01 | 88 | 2.6 | 4" | 0.000,006 |
| Flamsteed | Sex, Quad | 1725 | 1697 | 44 | 2934 | | 40" | |
| Lacaille, southern | Quad | (1763) | 1752 | 2 | 9766 | | 30" | |
| Mayer | Quad | 1775 | 1756 | 1.2 | 998 | | 6" | |
| Bradley/Auwers | TI+Quad | 1888 | 1760 | 12 | 3222 | | 1.1 | 0.002,7 |
| Lalande | Quad | 1801 | 1795 | 40 | 50,000 | | 4 | |
| Piazzi | AAC | 1814 | 1802 | 21 | 7646 | | 2.5 | |
| Lalande/Baily | compiled | 1847 | | | 47,390 | | | |
| Argelander | MC | 1867 | 1856 | 22 | 33,811 | 2 | 0.9 | 0.042 |
| Küstner | MC | 1908 | 1899 | 10 | 10,663 | 2.4 | 0.34 | 0.092 |
| USNO | MC | 1920 | 1907 | 8 | 4526 | 10 | (0.15) | 0.20 |
| USNO | MC | 1952 | 1945 | 8 | 5216 | 6 | (0.15) | 0.23 |
| Astrographic Cat. | Pgr. | | 1900 | 60 | 4,500,000 | 2 | 0.2 | 110 |
| Stoy | MC | 1968 | 1948 | 13 | 6800 | 2 | 0.43 | 0.037 |
| GC | MC | 1937 | 1900 | 175 | 33,342 | | 0.15 | |
| SAOC | Pgr.+MC | 1965 | 1930 | 50 | 259,000 | | 0.2 | |
| Perth70 | p.e. MC | 1976 | 1970 | 5 | 24,900 | 4 | 0.15 | 1.1 |
| FK5 | MC | 1988 | 1950 | 242 | 1535 | | 0.04 | |
| PPM N+S | Pgr.+MC | 1993 | 1945 | 90 | 379,000 | 6 | 0.04 | |
| CMC1-11 | p.e. MC | 1999 | 1991 | 14 | 176,591 | 6 | 0.07 | 36 |
| Hipparcos | p.e.sat. | 1997 | 1991 | 3 | 118,218 | 110 | 0.001 | 120,000 |
| Tycho-2 | p.e.sat. | 2000 | 1991 | 3 | 2,539,913 | 130 | 0.06 | 700 |
| USNO-B1.0 | Pgr. | 2002 | | 50 | 1,000,000,000 | | | |
| UCAC2 | CCD | 2003 | 2000 | 4 | 48,000,000 | 2 | 0.06 | 13,000 |
| 2MASS | HgCdTe | 2003 | 2000 | 3 | 400,000,000 | | 0.08 | 62,000 |
| CMC14 | CCD MC | 2005 | 2002 | 6 | 95,000,000 | 2 | 0.07 | 19,000 |
| GSC-II | Pgr. | 2005 | | 50 | 945,000,000 | | | |

**Photographic position catalogues:** The Astrographic Catalogue (AC) was obtained from photographic plates taken between 1892 and 1950, for details see Eichhorn (1974). It is the biggest astronomical enterprise ever undertaken by international cooperation and it began with a meeting in Paris in April 1887 invited by the French Academy of Science. It was a revolutionary idea at that time. Up to then, most star positions were obtained from observations on meridian circles. This imposed a limit of about 9[th] magnitude on stars accessible to observation; and (at that time) the typical standard error of a meridian position was about 0.5 arcsec in either coordinate. The photographic method then under consideration made it practical (without too much effort) to derive positions with a standard error about 0.3 arcsec for stars as faint as 13[th] or even 14[th] magnitude.

The plates were taken with identical telescopes at 20 observatories distributed at all geographic latitudes. The telescope, called a Normal Astrograph, had a lens of 33 cm aperture and 3.4 m focal length and a useful field of 2.1 x 2.1 square degrees. The plates were measured and star coordinates were published in about 150 volumes, the last ones in 1971. These books were inconvenient to use and the given coordinates suffered from the lack of an accurate reference system when the reductions were made. Due primarily to the enormous job of getting the data into machine readable form, attempts to attain a usable whole-sky catalogue failed until the 1990s. The USNO made a new reduction of the AC, containing over 4.5 million star positions, and published it in 2001 as AC 2000.2. The positions in AC 2000.2 and more than 140 other ground-based catalogues were used with the Tycho-2 positions to derive proper motions of the 2.5 million stars in the Tycho-2 Catalogue, as well as the UCAC2 (Zacharias 2008).

Other photographic astrometric enterprises with ever better lens astrographs giving a larger usable field were undertaken. They have resulted in hundreds of thousands of star positions but only some of the names can be mentioned: the AGK2 observed about 1930, published 1952, the AGK3 observed 1959-1961, published 1973, including proper motions, the Yale Catalogues observed 1914-1956, published in the 1950s, and the Cape Photographic Catalogue observed 1930-1953, published 1968. Much larger catalogues of a 1000 million stars were derived from Schmidt plates taken in the 1950s and later, going to much fainter magnitudes about 20[th]. From the USNO came the USNO A1.0, and B1.0 catalogues. The most recent Hubble Guide Star Catalogue is the GSC-II, obtained with the Palomar and UK Schmidt telescopes at two epochs and three photometric band passes. These catalogues and USNO B1.0 contain positions, proper motions, and photometry.

**CCD position catalogues:** After 1980 photographic plates were gradually replaced by electronic detectors, especially CCDs because of the higher sensitivity and the immediate digitization of the observation, and because astrometric quality plates were no longer manufactured. CCDs are primarily used in pointing mode. Scanning mode was introduced by Stone & Monet (1990) on a meridian circle. Scanning mode with CCDs on an astrometric satellite, Roemer, was proposed in 1992 by Høg (1993). This proposal initiated other similar projects, FAME and DIVA, and Roemer itself developed into the Gaia mission. From the web: The UCAC2 catalogue used the U.S. Naval Observatory Twin Astrograph of 20 cm aperture and a 4k by 4k CCD camera and it covers the declinations -90 to +40 deg. The catalogue positions have a standard error of 70 mas at the limiting magnitude of R=16, and an error of 20 mas at 11-14[th] mag. Such a smaller error for brighter stars is typical for many catalogues, but it is not expressed in the present tables. The 2MASS all-sky catalogue was obtained by two highly automatic telescopes with 1.3 m aperture equipped with HgCdTe detectors sensitive in the J,H,K bands (1-2 microns) with a limit of 17 mag in J. The CMC14 was obtained with the Carlsberg Meridian Circle with 18 cm aperture using CCDs in scanning mode giving a magnitude range of 9-17 in the red band, r'. The Tycho-2 Catalogue supplied the astrometric reference stars for UCAC2, CMC14 and GSC-II. Further projects for astrometry and photometry with enormous CCD arrays are: the ongoing Sloan Digital Sky Survey (SDSS), the coming Pan-STARRS, and the Gaia satellite.

**Table 2**  Proper motion catalogues.  The standard error, $s$, is a median value, if available. Mean epoch, observation period, number of catalogues, and the standard error are sometimes given as round numbers.  The observations used for the Tycho-2 proper motions were obtained by Hipparcos, MCs and photography, and the catalogues after 2000 benefit greatly from Tycho-2 as reference catalogue. - Abbr.: mas/yr = milliarcsecond/year

| Catalogue | Instrument | Publ. *year* | Mean epoch *year* | Obs. period *years* | N *entries* | n *catalogues* | s s.e. of star *mas/yr* | W weight | S_pos.publ. *mas* |
|---|---|---|---|---|---|---|---|---|---|
| Halley | | 1718 | | | 3 | | | | |
| Mayer | MC+Quad | 1775 | | | 80 | | | | |
| Mädler | Quad+MC | 1856 | | | 3222 | | | | |
| Auwers' FC | Quad+MC | 1879 | | 117 | 539 Dec > 10 | 9 | 8? | 8 | |
| NFK | Quad+MC | 1907 | | 155 | 925 | | 5? | 40 | |
| FK3 | MC | 1937 | 1900 | 190 | 1535 | 70 | 3 | 170 | 150 |
| GC | Quad+MC | 1937 | 1900 | 175 | 33,342 | 238 | 10 | 330 | 400 |
| N30 | MC | 1952 | 1930 | 100 | 5268 | 60 | 5 | 210 | 150 |
| SAOC | Pgr.+MC | 1965 | 1930 | 50 | 259,000 | 10 | 15 | 1200 | 560 |
| FK4 | MC | 1963 | 1920 | 213 | 1535 | 250 | 2 | 380 | 380 |
| FK5 | MC | 1988 | 1950 | 242 | 1535 | 350 | 1.2 | 1100 | 62 |
| PPM North | Pgr.+MC | 1991 | 1931 | 90 | 182,000 | 12 | 4.2 | 10,300 | 270 |
| PPM South | Pgr.+MC | 1993 | 1962 | 100 | 197,000 | 14 | 3.0 | 22,000 | 110 |
| PPM N+S | | | | | 379,000 | | 3.4 | 32,000 | 144 |
| Hipparcos | p.e.sat. | 1997 | 1991 | 3 | 118,218 | 1 | 0.9 | 120,000 | 6 |
| Tycho-2 | p.e.sat.++ | 2000 | 1991 | 100 | 2,430,468 | 145 | 2.5 | 400,000 | 64 |
| SPM3 | Pgr. | 2004 | 1980 | 23 | 10,700,000 | 2 | 4.0 | 670,000 | 100 |
| UCAC2 | CCD++ | 2003 | 1990 | 100 | 48,000,000 | 146 | 6.0 | 1,300,000 | 80 |
| USNO-B | Pgr. | 2002 | 1975 | 50 | 1,000,000,000 | 2 | 7.0 | 20,000,000 | 275 |

# Proper motion catalogues, Table 2

This small selection of proper motion catalogues shows especially the increasing number of stars and the improvement of accuracy over three centuries. All catalogues, apart from the first entries, contain both positions and proper motions, the motions being derived from positions observed over long periods of time and sometimes with different types of instruments. The number of catalogues, *n*, used for the proper motions is given, but this number should be regarded with caution since a single meridian circle catalogue and the entire AC are both counted as one catalogue although AC contributes much more weight to the derived proper motions.

Some of the catalogues in Table 2 are the especially accurate fundamental catalogues, Auwers' FC, NFK, N30, FK3, FK4 and FK5. They were compiled in order to provide reference stars for meridian circle observations and very accurate proper motions for the study of kinematics and dynamics of the Galactic stellar system. Scott (1963) gives an overview, including the proper motion errors for FK3 and N30. The FK5 states an error of 0.75 mas/yr, but from Tables 1 to 4 by Mignard & Froeschlé (2000) we have derived that the error is 1.6 times higher i.e. 1.2 mas/yr. A slightly larger value is given for FK4. The larger reference catalogues IRS and ACRS from the 1990s of respectively 36,027 and 320,211 stars for the reduction of photographic plates are described by T. Corbin in Høg (2008c).

Table 2 includes a standard error of the positions in the year of publication, $s_{pos.publ}$. This positional error is a measure for the ability of the catalogue to provide good positions in the years following the publication. The value is calculated by quadratic addition of the error due to proper motion and the position error at the mean epoch. The resulting error is mainly due to the proper motion error in all catalogues, except Tycho-2 where the 0.06 arcsec error at the mean epoch dominates by far. Tycho-2 will therefore keep its value as reference catalogue until 2016 when the first Gaia catalogue is released..

The large all-sky position and proper motion catalogues in the table of stars brighter than 11 mag are GC, SAOC, PPM, Hipparcos, and Tycho-2, published from 1937 to 2000. The progress from GC to Tycho-2 for the practical user of catalogues appears in the increase by a factor of almost 100 in the number of stars and more than 1000 in the weight. Furthermore, the errors of star positions at the time of publication of the catalogue decreased by a factor 6. After the year 2000 much larger catalogues with up to 1000 million stars cover also the fainter stars between 11 and 20 mag with positions, proper motions and multi-colour photometry.

## Catalogues of trigonometric parallaxes, Table 3

The first three reliable annual parallaxes of stars were published in 1838-40, their standard errors are given according to Høg (2008c). The technology and methodology of parallax measurement (see the table and Høg 2008c,d) remained basically unchanged for about 60 years (ca. 1840-1900). Despite the rapid initial success, the number of stars with reliable parallaxes grew slowly, and is hard to calculate because of disputes about which were reliable. The uncertainties and systematic errors are strikingly illuminated when we see that the review paper with catalogue by Oudemans from 1889 lists 55 observations of 61 Cyg and gives the mean parallax as 0.40". This is 0.11" larger than the true value, a deviation 8 (eight) times larger than the formal error of 0.014" claimed by Bessel in 1840, and Bessel's own parallax deviates four times his own error from the true value. In 1899 Ch. Andre gives the parallax as 0.44" from the same 55 observations, see references and discussion in Høg (2008c).

A catalogue by Bigourdan (1909) lists trigonometric parallaxes for about 300 stars, a few with up to 40 observations. The consistency of multiple observations indicates a precision (i.e. internal error, therefore in brackets) about 50 mas per observation, and a median precision of 30 mas may be inferred for about 200 stars having more than one observation. Many observations are shown (by bold face) to be the average of several measurements by the same observer, including most of the 100 with only one observation. Russell (1910) presents 52 new photographic parallaxes and claims a standard error about 40 mas.

**Table 3** Catalogues of trigonometric parallaxes.

| Catalogue | Instrument | Publ. year | Obs. period years | N entries | s s.e. of star mas | W weight | Notes |
|-----------|-----------|------------|-------------------|-----------|--------------------|----------|-------|
| Bessel | Heliometer | 1838 | | 1 | 60 | | |
| Henderson | Quad | 1839 | | 1 | 500 | | |
| Struve | Wire micr. | 1840 | | 1 | 100 | | |
| Peters | Visual | 1850 | | 20 | ? | | |
| Oudemans | Visual+Pgr. | 1889 | 60 | 50 | ? | | |
| Bigourdan | Visual+Pgr. | 1909 | | 100 | (50) | 0.04 | With one observation per star |
| - same - | | | | 200 | (30) | 0.2 | With two or more obs. per star |
| Russell | Pgr. | 1910 | | 52 | (40) | 0.03 | |
| Schlesinger | Pgr. | 1935 | 35 | 7534 | 15 | | Includes spectroscopic par. |
| Jenkins | Pgr. | 1952 | 50 | 5822 | 15 | 26 | |
| Van Altena | Pgr. | 1995 | 95 | 8112 | 10 | 81 | |
| - same - | | | | 1649 | | | Error of parallax < 17.5 % |
| - same - | | | | 940 | | | Error of parallax < 10 % |
| Hipparcos | p.e.sat. | 1997 | 3 | 118,218 | 1.0 | 120,000 | |
| - same - | | | | 20,853 | | | Error of parallax < 10 % |
| USNO | Pgr.+CCD | -2008 | 20 | 357 | 0.6 | 1000 | C. Dahn, priv. comm. 2008 |
| HST | CCD, satellite | -2008 | 18 | 31 | 0.24 | 500 | F. Benedict, priv. comm. 2008 |

For the 1952 parallax catalogue by Jenkins a standard error of 15 mas is derived by Hertzsprung (1952). A round value of 10 mas is given for the median standard error of the last ground-based catalogue (van Altena 1995), but the errors vary greatly, viz. between 1 and 20 mas.

Hipparcos obtained a median standard error of 1.0 mas for parallaxes. A similar or better accuracy has been achieved from the ground and with the Hubble Space Telescope (HST) for several hundred much fainter stars. The number of parallaxes with an error less than a given fraction of the parallax value is given in three cases.

**Acknowledgements:** I am indebted to Adriaan Blaauw for the kind invitation to contribute to the symposium in 2008 on this vast subject. Without the invitation I would never have engaged myself in this quite large undertaking. Comments to previous versions of the paper from F. Arenou, P. Brosche, A. Chapman, T. Corbin, D.W. Evans, C. Fabricius, J. Lequeux, F. Mignard, H. Pedersen, A. Schrimpf, P.K. Seidelmann, C. Turon, S.E. Urban, W.F. van Altena, R. van Gent, F. Verbunt, A. Wittmann and N. Zacharias are gratefully acknowledged.

# References

*The following list includes some of the references in Høg (2008d).* *New references since 2008 are red.*